\documentstyle[aps,prb,graphics,twocolumn]{revtex}

\newcommand{\tpf}{(TMTTF)$_2$PF$_6$ }

\begin{document}

\title{From spin--Peierls to superconductivity: (TMTTF)$_2$PF$_6$ under high
  pressure}

\author{D. Jaccard and H. Wilhelm}

\address{DPMC, Universit\'e de Gen\`eve, Quai E.--Ansermet 24, 1211
Geneva 4,
  Switzerland}

\author{D. J\'{e}rome and J. Moser} 

\address{Laboratoire de Physique des Solides (associ\'{e} au CNRS),
  Universit\'{e} Paris--Sud, 91405 Orsay, France}

\author{C. Carcel and J. M. Fabre}

\address{Laboratoire H\'{e}t\'{e}rochimie et Mat\'eriaux Organiques ENSCM/ESA--5076,
  34296 Montpellier Cedex 5, France} 

\date{\today}

\maketitle

\widetext
\begin{abstract}
  The nature of the attractive electron--electron interaction, leading to the
  formation of Cooper--pairs in unconventional superconductors has still to be
  fully understood and is subject to intensive research.  Here we show that
  the sequence spin-Peierls, antiferromagnetism, superconductivity observed in
  \tpf under pressure makes the (TM)$_2$X phase diagram universal. We argue
  that the suppression of the spin--Peierls transition under pressure, the
  close vicinity of antiferromagnetic and superconducting phases at high
  pressure as well as the existence of critical antiferromagnetic fluctuations
  above $T_c$ strongly support the intriguing possibility that the interchain
  exchange of antiferromagnetic fluctuations provides the pairing mechanism
  required for bound charge carriers.
\end{abstract}
 
\narrowtext
\tighten

\vspace*{4mm} 
The existence of a common border between the superconducting (SC) ground state
and the insulating phase of spin density wave (SDW) nature \cite{Jerome82},
was recognized as a remarkable property of the phase diagram of the Bechgaard
salt (TMTSF)$_2$PF$_6$. It belongs to a broad family of isostructural
compounds (TM)$_2$X, where the flat organic molecule TM is either
tetramethyltetraselenafulvalene (TMTSF) or tetramethyltetrathiafulvalene
(TMTTF). Here X denotes a monovalent anion such as PF$_6$, AsF$_6$, ClO$_4$ or
Br \cite{Ishiguro98}. In the crystal, these molecules form stacks separated by
chains of anions X. The overlap between the electron clouds of neighboring TM
molecules along the stacking direction ($\parallel$ the $a$--axis) is about 10
(500) times larger than that between the stacks in the transverse
$b$--($c-$)direction.  Provided that the longitudinal overlap is large
compared to the on--site Coulomb repulsion, these organic materials become
conducting with a pronounced one dimensional (1--D) character.

The 1--D character of the Fermi surface of (TMTSF)$_2$ PF$_6$, the presence of
a spin-Peierls (SP) transition instead of the usual Peierls instability
\cite{Pouget82,Ribault80} as well as the existence of enhanced
antiferromagnetic (AF) fluctuations at low temperature, evidenced by NMR
relaxation experiments, raised several questions about the mechanism
responsible for superconductivity in organic conductors \cite{Creuzet87}.
Since 1--D physics is a relevant concept in these low dimensional systems, SDW
and electron--electron pairing can develop simultaneously at low temperature
in the interacting electron gas \cite{Jerome82}. A cross--over from SDW to SC
correlations could possibly be achieved through a small variation of the
coupling constants either by applying pressure or changing X\cite{Emery83}.
Furthermore, the nuclear spin--lattice relaxation rate data of
(TMTSF)$_2$PF$_6$ suggest that SDW correlations prevail at low temperature
even under pressure when superconductivity is stabilized \cite{Creuzet87}.

In the generic phase diagram proposed for the (TM)$_{2}$X 
\newpage
\vspace*{31mm}
\noindent
family \cite{Jerome91} the sequence of ground states (SP, AF/SDW and SC) can be
observed for different members of the series if they are placed according to
their ambient pressure properties.  Even parts of the sequence can be found
for a given member of the series if pressure is applied. For instance, the SDW
ground state of (TMTTF)$_2$Br can be suppressed and at a pressure $P= 2.6$~GPa
a SC phase appears \cite{Balicas94}.  However, starting from a SP ground
state, which is observed only for (TMTTF)$_2$X with X=PF$_6$ or AsF$_6$
\cite{Pouget82,Brun77,Coulon82}, no superconductivity had been observed.

In this context \tpf is of particular interest because the existence of the
pressure--dependent SP ordering can be used for a quantitative estimate of the
pressure dependence of the pairing force, mediated by acoustic phonons
\cite{Bourbonnais95}. The coupling between electrons and the lattice manifests
itself by a divergence of the $2k_{\rm F}$ lattice susceptibility below 100 K
\cite{Pouget82} and by opening a pseudo--gap in the uniform spin
susceptibility at $T^{0}_{\rm SP} \approx 40$~K \cite{Creuzet87b}. The latter
evolves towards a true SP gap at $T_{\rm SP} = 19$~K \cite{Pouget82}. When
1--D $2k_{\rm F}$ phonon softening occurs in the presence of fully developed
1--D AF correlations in the Mott localized phase at $T < T_{\rho}$ (where
$T_{\rho}$ represents the temperature below which the 1--D Mott localization
produces an insulating behavior of the electrical resistivity) bond charge
correlations couple to $2k_{\rm F}$ acoustic phonons and a SP instability sets
in at low temperature \cite{Bourbonnais95}. The SP ordering is suppressed
under pressure and a N\'{e}el state is stabilized above $P=0.9$~GPa
\cite{Creuzet87,Chow98}.  As the SP instability involves the electron-phonon
interaction it can be inferred from the experimentally determined pressure
dependence of $T_{\rm SP}$ in the low pressure regime that the bare
electron-acoustic phonon interaction should be severely depressed at higher
pressure \cite{Bourbonnais95}. A small value of the electron--phonon
coupling in addition to the weakness of the $2k_{\rm F}$ lattice
susceptibility in (TM)$_2$X compounds whenever a SP order is {\it not} the
ground state \cite{Pouget97} would preclude the stabilization of a SC ground
state above the millikelvin range. Consequently, a relevant question to be
raised is whether superconductivity can also be stabilized under pressure in
(TMTTF)$_2$PF$_6$ as it is in its selenide based analog (TMTSF)$_2$PF$_6$.
This has been a great stimulus for the search and subsequent discovery of
superconductivity in (TMTTF)$_2$PF$_6$ at pressures beyond 4 GPa.

%
%

\begin{figure}[h]
    \caption{The sample chamber of the high pressure device before
      pressurization. In the cylindrical gasket (pyrophyllite) the \tpf single
      crystal (black bar) is placed on a disk of a soft pressure medium
      (steatite, $\phi=2$mm). Thin Au--wires ($\phi=5\mu$m) on the sample and
      the pressure gauge (Pb--foil) are attached to thicker Au--wires
      ($\phi=50\mu$m) which establish the electrical contact across the
      gasket. The cell is closed with a second disk of steatite on top of this
      arrangement and then pressurized between the two Bridgman anvils. }
    \label{fig:cellgeneva}
\end{figure}

The influence of pressure on the longitudinal electrical resistivity $\rho_a$
of (TMTTF)$_2$PF$_6$ single crystals grown by
electrocrystallization\cite{Mora92} was studied with a piston--cylinder
clamped cell capable to reach pressures of $P_{\rm max}\approx 4$~GPa and a
Bridgman anvil cell for higher pressures ($P_{\rm max}\approx 10$~GPa)
designed for temperatures as low as $T=25$~mK \cite{Jaccard98}. The sample
chamber of the latter apparatus is shown in Fig.~\ref{fig:cellgeneva} \cite{grad}.  
 
The pressure effect on $\rho_a$ at room temperature is quite strong: $\rho_a$
decreases from 400~m$\Omega$cm at ambient pressure to $\rho_a=11$~m$\Omega$cm
for $P=4.05$~GPa. At this pressure, temperature has a similar strong influence
on $\rho_a(T)$: it decreases by a factor of 40 upon cooling to temperatures of
the order of 10~K. Here, $\rho_a(T)$ passes through a minimum at $T_{\rm min}$
and the upturn in $\rho_a(T)$ is related to a transition into an insulating
state, attributed to the onset of itinerant antiferromagnetism (SDW)
\cite{Moser98}. 
Beyond $P=4$~GPa the strong increase of $\rho(T)$ at low
temperature is disrupted by the onset of a sharp drop in resistivity at $T_c =
1.8$~K (see curve at $P=4.35$~GPa in Fig.~\ref{fig:rhogeneva}).  
At $P=4.73$~GPa $\rho(T)$ already starts to decline at $T_c=2.2$~K and has
decreased by one order of magnitude at 1~K. The temperature $T_c$ as well as
the magnitude of the drop in resistivity decrease as pressure increases
further.  The residual resistivity $\rho_0$, measured at the lowest
temperature reached in each run, amounts to 1--2~m$\Omega$cm.  Beyond 7~GPa no
evidence of a drop in resistivity is found above 50 mK.

%
%

\begin{figure}[]
    \caption{Electrical resistivity $\rho(T)$ of \tpf at various pressures and
      low temperature.}
    \label{fig:rhogeneva} 
\end{figure}

The influence of an external magnetic field along the $c$--axis is shown in
Fig.~\ref{fig:rhosouschamp}. The drop in resistivity is completely suppressed
in a field of $\mu_0H=0.8$~T. This is taken as a strong argument to identify
$T_c$ as a SC transition temperature despite the finite value of the
residual resistivity which can be attributed (above 4.7 GPa) to microcracks
related to the extreme sample brittleness and possible non--hydrostatic
components in the Bridgman anvil cell.  The value of the critical field
$H^{c}_{c2}$, determined by the recovery of the normal state resistivity,
increases together with $T_c$ as pressure decreases. Within the framework of
clean type II superconductors which is justified in most (TMTSF)$_2$X salts
since the electron mean free path is of the order of $10^4\times a$, with $a$
the lattice parameter, $dH^{c}_{c2}/dT$ = $-\frac{A}{t_{a}t_{b}} T_c$ for
$T\rightarrow T_c$, where $A$ is a constant independent of the field
orientation and pressure\cite{Gorkov85}. The variation of
$dH^{c}_{c2}/T_{c}dT$ between 4.45 and 6.14~GPa leads to a pressure dependence
of $2.0$\%/kbar for $(t_{a}t_{b})^{1/2}$ which is in fair agreement with the
optical measurements of the bare band parameters of organic conductors under
pressure\cite{Jerome82}.
 
Our data provide the missing information enabling the ($T$,$P$) phase diagram
of (TMTTF)$_2$PF$_6$ shown in Fig.~\ref{fig:phasediagram} to be constructed
and establish its truly universal character. After the suppression of the SP
phase \cite{Creuzet87,Chow98} the AF (N\'eel and SDW) ground states are stable
up to about 4~GPa \cite{Moser98}.  The SC region extends from slightly above
4~GPa to almost 7~GPa with a SC transition temperature as high as $T_c=2.2$~K
at $P=4.73$~GPa.  It is worth noting that close to this pressure three phase
lines meet. The green region in Fig.~\ref{fig:phasediagram} indicates the
presence of AF spin fluctuations. This region has an upper bound in
temperature defined by $T_{\rm min}$ and a lower limit given by either $T_{\rm
  SDW}$ or $T_c$. In this temperature interval $\rho(T)$ shows an upturn. The
width in temperature of this interval increases with decreasing pressure and
is largest where $T_c(P)$ reaches its optimum value.  Thus, $T_{\rm min}$
appears to be closely linked to the critical temperature $T_c$. Critical AF
fluctuations seem to be enhanced when the SDW ground state is approached from
high pressure, i.~e., where the system is close to the border between the SDW
and SC phases. At slightly lower pressure the decrease of $T_c$ is clearly
related to the occurrence of the SDW phase at a higher temperature.  Similar
behavior is encountered in the competition between charge density wave and SC
instabilities \cite{Friedel75}. The correlation between the fall of $T_{\rm
  SDW}$ and the rise in $T_c$ reflects the suppression of the SDW with
pressure. This restores areas of the Fermi surface lost by the creation of
magnetic gaps, thereby increasing the density of states at the Fermi level and
hence $T_c$.

%
%

\begin{figure}[t]
    \caption{Magnetic field dependence of $\rho(T)$ at $P=4.73$ GPa for 
      a field direction along the crystallographic $c$--axis. In zero field
      the critical temperature is as high as $T_c=2.2$~K. The inset shows
      $\partial H_{c2}/T_c\partial T_c$ vs pressure. }
    \label{fig:rhosouschamp} 
\end{figure}

The reentrance of superconductivity below the SDW ordering appears to be a
general behavior among (TM)$_2$X superconductors. It has also been identified
with a finite value of the electrical resistivity in the ``superconducting"
state in (TMTSF)$_2$AsF$_6$ \cite{Jerome82,Azevedo84} although less clear due
to a larger compressibility. The enhanced $\rho_0$--values reported here
cannot be attributed to pressure inhomogeneity. They support furthermore the
picture of an inhomogeneous SC ground state between 4.2 and 4.5~GPa where SC
islands are dispersed in a SDW insulating background. In such a scenario the
non--percolating SC domains could contribute to a finite resistivity below
$T_c$.

%
%

\begin{figure}[t]
\vspace*{5mm}
    \caption{($T$,$P$) phase diagram of (TMTTF)$_2$PF$_6$. The spin Peierls
      (SP), an antiferromagnetic (AF), and the spin density wave
      state (SDW) are suppressed by pressure and a superconducting (SC) phase
      emerges above 4~GPa. Over a wide pressure range AF spin fluctuations
      (green region) are present. At high temperature a Mott--Hubbard
      insulating (M--H I) and a metallic (M) state are observed. Open symbols
      represent data taken from Ref.~$[18]$. }
    \label{fig:phasediagram} 
\end{figure}

Given the inability for the traditional electron--phonon mechanism to promote
SC in (TMTTF)$_2$PF$_6$ under very high pressure in the range of 2~K other
approaches such as magnetically mediated pairing can be considered. The
correlation between the $T_c$--value and the (insulating) spin fluctuation
regime is taken as a strong experimental argument in favor of a pairing
mechanism involving AF fluctuations. Such a scenario was also considered for
Ce-- and U--based strongly correlated electron multiband systems where
deviations from the canonical quadratic temperature dependence of the Fermi
liquid model are observed at the border between superconductivity and
antiferromagnetism \cite{Julian96,Mathur98}.

The use of SDW fluctuations to form bound states of charge carriers is another
possibility considered by Emery and worked out in the context of nearly AF
itinerant fermion systems \cite{Beal86}. As far as 1--D organic conductors are
concerned, it was shown that the exchange of SDW fluctuations between carriers
belonging to the same stack does not lead to attractive pairing and thus the
development of a 1--D attractive pairing appears to be hopeless. In addition,
in 1--D systems the electron--phonon coupling is opposed to the Coulomb
repulsion of carriers moving in a restricted phase space.

A conventional approach using the spin fluctuation exchange model in
quasi--1--D organic superconductors has predicted d--wave pairing in the
vicinity of the SDW phase \cite{Kino99}. However, this theory does not take
fully into account the entire temperature regime and in particular the
non--Fermi liquid features, observed in DC transport \cite{Moser98,Danner95}
and optical conductivity \cite{Vescoli98} at high temperature, which persist
down to low temperature. An attractive interstack pairing can be the
outcome of the exchange of AF spin fluctuations between electrons located on
neighboring stacks \cite{Bourbonnais88}. Such a mechanism would lead to an
anisotropic SC gap. In this picture, the attractive interaction is accounted
for by the growth of electron--hole pair tunneling generated in the 1--D
regime at temperatures larger than the 1--D to 2--D dimensionality cross--over
\cite{Bourbonnais95}.

To summarize, we have reported pressure--induced superconductivity in
(TMTTF)$_2$PF$_6$ {\it in spite of} its SP ground state at ambient pressure.
This important result establishes the universality of the (TM)$_2$X phase
diagram. Furthermore, the suppression of the SP ground state makes the
traditional phonon--mediated Cooper--pair formation unlikely to explain the
existence of superconductivity at temperatures as high as $T_c=2.2$~K at
$P=4.7$~GPa \cite{Adachi00}. The manifestation of critical AF spin
fluctuations above the onset of superconductivity and the close connection
between their amplitude and the value of $T_c$ speaks strongly in favor of an
interstack pairing mechanism mediated by the exchange of these fluctuations
between neighboring stacks. Thus, our findings supply an important input for
theoretical models of magnetic coupling in quasi 1--D conductors and may even
shed light on superconductivity in strongly correlated electron systems
including high--temperature superconductors.

Useful discussions with C. Bourbonnais, C. Pasquier, and P.  Auban--Senzier as
well as the technical help of M. Nardone are acknowledged.  The work carried
out in Geneva was supported by the Swiss National Science Foundation. We thank
R. Cartoni and A. Holmes for technical assistance.

%
%


\begin{references}
\bibitem{grad} The pressure dependent SC transition temperature of Pb yields
  the pressure. The upper limit for the inevitable, but small pressure gradient
  is estimated from the width of the Pb transition to be about 0.1~GPa.
\end{references}
\end{document}